\documentclass[12pt]{iopart}
\usepackage{amssymb}

\usepackage{graphicx} 
\usepackage[dvipsnames]{xcolor}

\expandafter\let\csname equation*\endcsname\relax
\expandafter\let\csname endequation*\endcsname\relax
\usepackage{amsmath}

\newcommand{\argmin}{\mathop{\mathrm{argmin}}\limits}
\usepackage{booktabs}

\usepackage[hyphens]{url}
\usepackage[hidelinks, urlcolor=cyan]{hyperref}  
\usepackage[hyphenbreaks]{breakurl}

\usepackage{comment}

\usepackage{subcaption}
\usepackage{tabularx}
\usepackage{booktabs}
\usepackage{caption} 

\usepackage{orcidlink}
\usepackage[normalem]{ulem} 
\usepackage{multirow}
\usepackage{bm} 

\usepackage[sorting=nyt,style=apa,doi=false,url=false,firstinits=true,terseinits=true]{biblatex}
\renewrobustcmd*{\bibinitdelim}{\,} 

\begin{document}
\citetrackerfalse

\title[]{Reduction of motion artifacts from photoplethysmography signals using learned convolutional sparse coding}

\author{Giulio Basso$^{1,3,*}$\orcidlink{0000-0002-4532-4354}, Xi Long$^{1,3,*}$\orcidlink{0000-0001-9505-1270}, Reinder Haakma$^{2,3}$\orcidlink{0000-0002-7245-7072}, Rik Vullings$^{1,3}$\orcidlink{0000-0002-2392-6098}}

\address{$^1$ Department of Electrical Engineering, Eindhoven University of Technology, Eindhoven, The Netherlands}
\address{$^2$ Department of Patient Care \& Monitoring, Philips Research, Eindhoven, The Netherlands}
\address{$^3$ Eindhoven MedTech Innovation Center (e/MTIC), Eindhoven, The Netherlands}
\begin{indented}
\item[] * Author to whom any correspondence should be addressed 
\end{indented}
\ead{g.b.basso@tue.nl}
\vspace{10pt}

\begin{abstract}
\emph{Objective.} Wearable devices with embedded photoplethysmography (PPG) enable continuous non-invasive monitoring of cardiac activity, offering a promising strategy to reduce the global burden of cardiovascular diseases. However, monitoring during daily life introduces motion artifacts that can compromise the signals. Traditional signal decomposition techniques often fail with severe artifacts. Deep learning denoisers are more effective but have poorer interpretability, which is critical for clinical acceptance. This study proposes a framework that combines the advantages of both signal decomposition and deep learning approaches. \emph{Approach.} We leverage algorithm unfolding to integrate prior knowledge about the PPG structure into a deep neural network, improving its interpretability. A learned convolutional sparse coding model encodes the signal into a sparse representation using a learned dictionary of kernels that capture recurrent morphological patterns. The network is trained for denoising using the PulseDB dataset and a synthetic motion artifact model from the literature. Performance is benchmarked with PPG during daily activities using the PPG-DaLiA dataset and compared with two reference deep learning methods. \emph{Main results.} On the synthetic dataset, the proposed method, on average, improved the signal-to-noise ratio (SNR) from –7.07 dB to 11.23 dB and reduced the heart rate mean absolute error (MAE) by 55\%. On the PPG-DaLiA dataset, the MAE decreased by 23\%. The proposed method obtained higher SNR and comparable MAE to the reference methods. \emph{Significance.} Our method effectively enhances the quality of PPG signals from wearable devices and enables the extraction of meaningful waveform features, which may inspire innovative tools for monitoring cardiovascular diseases.
\end{abstract}

%
\vspace{2pc}

\section{Introduction} \label{sec_introduction}

Photoplethysmography (PPG) is an optical technique that measures blood volume variations in the microvascular bed of skin \parencite{allen2007photoplethysmography}. Since PPG can be embedded in smartwatches and wristbands commonly used in daily life, it represents a non-invasive and low-cost solution to monitor the cardiovascular activity of patients. In addition, these wearable devices can acquire data continuously, increasing the chance of detecting critical pathological events and potentially enabling the early diagnosis of cardiovascular diseases.

Nevertheless, monitoring patients during daily activities inevitably introduces motion artifacts that can compromise the quality of the signals. 
These artifacts can be either periodic or non-periodic and often have a much larger amplitude than the pulsatile component of PPG \parencite{zhang2019motion}. Furthermore, the frequency band of the artifacts and the pulsatile PPG component often overlap \parencite{lee2007periodic}. Hence, effective denoising is challenging, especially when a motion reference signal from an accelerometer sensor is not available.


Denoising techniques that do not need a reference motion signal are often based on traditional signal decomposition techniques such as wavelet thresholding \parencite{lee2003reduction} \parencite{raghuram2010evaluation} \parencite{biswas2019motion} \parencite{peng2015comb}. The main assumption is that many real-world signals become sparse in the wavelet domain, with a few large coefficients representing the signal, while small coefficients are typically associated with noise and can be removed via thresholding. A key limitation is that the choice of an appropriate threshold and basis function, i.e. the mother wavelet, has a critical impact on the performance of these methods \parencite{han2017electrocardiogram}. For that reason, empirical mode decomposition (EMD), which extracts the basis functions from the signal itself, has been adopted for PPG denoising \parencite{wang2010artifact} \parencite{sun2012robust} \parencite{tang2016ppg}.

Signal decomposition techniques are valuable for their transparency, as they rely on well-defined signal modelling assumptions that facilitate the interpretation of how artifacts and target signals are distinguished. Nevertheless, these methods were demonstrated to fail in scenarios with high-intensity motion artifacts \parencite{wang2022ppg} \parencite{zhang2014troika}. This might be because the pulsatile PPG component and the motion artifacts can have similar time-frequency characteristics, especially when the artifacts are caused by periodic or large movements. As a result, signal decomposition methods like wavelet denoising or EMD may not effectively separate the two components or can inadvertently suppress relevant physiological information. Indeed, in both \parencite{lee2018bidirectional} and \parencite{mohagheghian2023noise}, applying wavelet or EMD denoising before heart rate (HR) estimation led to a less accurate estimation than without using any denoiser. 

Alternatively, deep learning techniques have been employed for denoising PPG signals and proved to outperform traditional signal decomposition methods \parencite{lee2018bidirectional} \parencite{wang2022ppg} \parencite{mohagheghian2023noise} \parencite{jain2023self}. One popular approach is represented by denoising autoencoders \parencite{lee2018bidirectional} \parencite{jain2023self} \parencite{mohagheghian2023noise}. In particular, \textcite{lee2018bidirectional} introduced a bidirectional recurrent denoising autoencoder (BRDAE), which uses bidirectional long-short term memory cells to leverage the sequential structure of the PPG signal. In \parencite{mohagheghian2023noise}, they proposed a convolutional denoising autoencoder (CDA) with skip connections that process the short-time Fourier transform (STFT) of the PPG signals, and demonstrated that this method outperforms the BRDAE method. Motivated by the recent popularity of generative models such as generative adversarial networks (GANs) and diffusion models, many studies employed generative models for PPG denoising \parencite{zheng2022ppg} \parencite{wang2022ppg} \parencite{afandizadeh2023accurate} \parencite{long2023reconstruction} \parencite{avila2025generative} \parencite{xia2025effective}. In \parencite{afandizadeh2023accurate}, they transformed the noisy PPG signals into 2D images and reconstructed the clean PPG signals using a cycle generative adversarial network (CycleGAN). \textcite{wang2022ppg} proposed a GAN inspired by GANomaly \parencite{akcay2019ganomaly} that is trained exclusively on clean PPG signals and learns to predict the succeeding clean PPG segment given the current one. Denoising is performed during deployment by taking a noisy segment and feeding the previous clean segment to the GAN to predict the current clean segment. Another method adapted from GANomaly is the FC-GAN proposed by \textcite{avila2025generative}, which replaces the convolutional layers with fully connected ones. Unlike the method in \parencite{wang2022ppg}, the denoiser in \parencite{avila2025generative} is trained using PPG signals corrupted with synthetic motion artifacts as input and the corresponding clean signals as targets. This approach outperformed the CycleGAN and the GAN by \textcite{wang2022ppg}.

However, the aforementioned deep learning approaches do not have the same transparency and interpretability as the signal decomposition methods. Interpretability is defined as the ability to understand the sequence of operations of a learned model \parencite{li2021deep}. By contrast, GANs and convolutional autoencoders behave as black boxes because they are purely data-driven, as they do not include any prior signal modelling assumptions. The model interpretability is crucial in the medical domain to ensure that experts have a clear understanding of the algorithms and trust their outcomes \parencite{min2017deep}. Secondly, more transparent methods facilitate troubleshooting, as it is easier to get insights into their underlying mechanisms and identify potential issues when the methods do not perform as expected.

In their seminal work, \textcite{gregor2010learning} introduced algorithm unfolding, which is a paradigm used to design neural networks by mimicking the iterations of a model-based iterative algorithm. One of the main advantages is that algorithm unfolding improves the model interpretability because the network inherits prior knowledge about the signal structure rather than learning it from intensive training data \parencite{li2021deep}. Additionally, unfolded networks are computationally faster and more data-efficient than purely data-driven networks, as they require less training data \parencite{li2021deep}.

Based on the reasons previously mentioned, this study investigates a deep learning framework that combines the advantages of signal decomposition and deep learning approaches. This is done by deriving the network from a prior sparse signal decomposition model using algorithm unfolding.
The resulting model consists of an encoder based on the deep unfolded iterative shrinkage algorithm \parencite{van2019deep} followed by a single-convolutional layer decoder \parencite{sreter2018learned}. The model was trained using a synthetic dataset based on the PulseDB dataset \parencite{wang2023pulsedb} and corrupted with synthetic motion artifacts \parencite{paliakaite2021modeling}. The method was then benchmarked on PPG signals during daily life activities using the PPG-DaLiA dataset \parencite{reiss2019deep} and compared to two reference methods \parencite{mohagheghian2023noise} \parencite{avila2025generative}.

\section{Methods} \label{label_sec2}
\subsection{Datasets and preprocessing} \label{label_sec2.1}
In this study, two datasets were used: a synthetic dataset based on the PulseDB dataset \parencite{wang2023pulsedb} to train the proposed method and the PPG-DaLiA dataset \parencite{reiss2019deep} to test the proposed method with motion artifacts from a real-life scenario.

PulseDB is a large, cleaned dataset consisting of fingertip PPG, ECG and arterial blood pressure waveforms recorded from intensive care units (ICUs). The dataset comprises a total of 14,570 hours of recordings from 5,361 subjects. The waveforms were extracted from the MIMIC-III waveform database matched subset \parencite{MIMICIII}\parencite{johnson2016mimic}\parencite{goldberger2000physiobank} and the VitalDB waveform database \parencite{lee2022vitaldb} using an extensive data cleaning procedure. All signals are provided at a sampling rate of 125 Hz and divided into 10-second segments.

We created a synthetic dataset using PPG signals recorded in ICUs from the PulseDB dataset, which were considered as the clean references. This assumption is based on the fact that ICU recordings are typically less affected by motion artifacts, since ICU patients are generally sedated or have limited mobility.
As suggested in \parencite{wang2023pulsedb}, we selected the first 400 10-second segments per subject to ensure equal contribution from each subject and balance the number of subjects from the MIMIC-III matched subset and the VitalDB database.
Then, the subjects were randomly split into training, validation and test sets with proportions of approximately 70\% (2024 subjects), 15\% (434 subjects) and 15\% (436 subjects), respectively.

Afterwards, clean segments from the PulseDB dataset were corrupted with synthetic motion artifacts using the model from \parencite{paliakaite2021modeling}\footnote{The implementation of the synthetic motion artifact model is available at \url{https://doi.org/10.13026/s32e-sv15}}, such that each corrupted segment had its corresponding clean target segment for training our network. A probabilistic model was developed in \parencite{paliakaite2021modeling} by characterizing the artifacts from PPG signals acquired during cardiac rehabilitation. The artifacts were divided into four types, namely device
displacement, forearm motion, hand motion, and poor contact. Each type of artifact was characterized based on its amplitude, spectral slope, duration and transition probability, and probability distributions were fitted to each of these features. Then, synthetic artifacts were generated by filtering white noise with an FIR filter whose characteristic features were sampled by the probability distributions previously fitted. Finally, a continuous-time Markov chain was used to model the transitions between artifact-free and artifacts-corrupted intervals.

In our synthetic dataset, we did not model the probability of transitions between artifact-free and artifacts-corrupted intervals using the Markov chain, because we rather simulated one single artifact per PPG segment. The durations of the artifacts $T_{MA}$ were sampled from a uniform distribution with an interval of [1, 10] seconds. The starting instants of the artifacts were sampled from a uniform distribution between [0, $10-T_{MA}$] seconds. All the segments of one subject had artifacts of the same type, and we ensured the same number of subjects for each type of artifact. 

Both clean and artifact-corrupted PPG signals from the synthetic dataset were filtered with a 4th-order Chebyshev type II band-pass filter between [0.5, 18] Hz. Finally, each PPG segment was normalized in amplitude between 0 and 1.

The second dataset used in this study was PPG-DaLiA, which includes wrist PPG, ECG, and accelerometer data from 15 subjects during daily life activities, spanning approximately 36 hours of data. The subjects performed various activities (sitting still, ascending/descending stairs, table soccer, cycling, driving car, lunch break, walking and working), which ensured very diverse motion artifacts and also a broad range of HR. Between each activity, there was a transient period that included the time to get to the starting point and prepare for the next activity. Wrist PPG signals are provided with a sampling rate of 64 Hz. Furthermore, the dataset includes ground-truth instantaneous HR estimated from ECG every 8-second window with a 2-second shift. To be consistent with the synthetic dataset, we upsampled the signals from PPG-DaLiA to 125 Hz and applied the same preprocessing steps.

\subsection{The proposed method} \label{section_2.2}

\begin{figure}
    \centering
    \includegraphics[width=\linewidth]{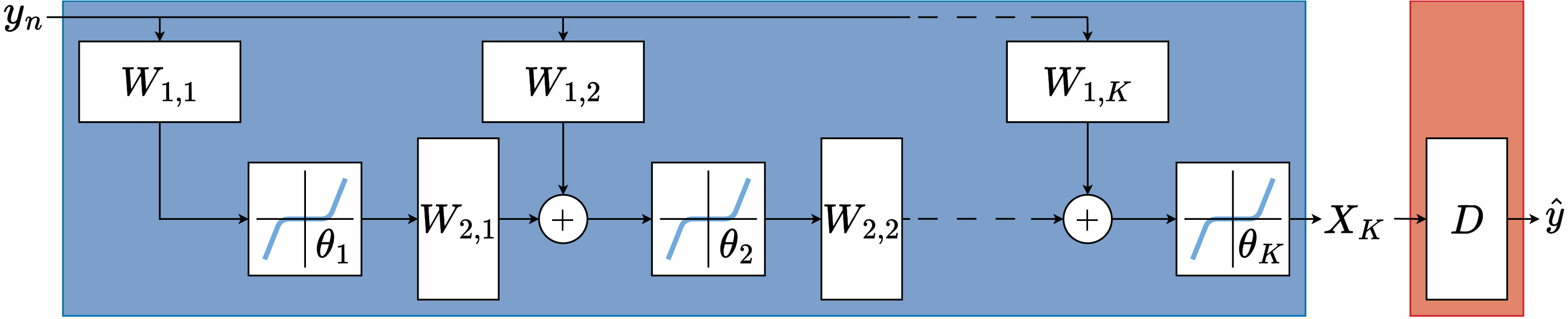}
    \caption{Architecture of the proposed method. $y_n$ represents the corrupted PPG signal which forms the input, $X_K$ is the sparse code and $\hat{y}$ is the reconstructed PPG signal. The deep unfolded iterative shrinkage algorithm encoder is highlighted in light blue. $K$ is the number of folds, $W_{1,k}$ and $W_{2,k}$ are the convolutional layers of the $k$-th fold (equation \ref{label_eq2}), while $\theta_k$ are the learned thresholds of the smooth soft-thresholding functions. The single convolutional layer decoder is highlighted in red, where $D$ represents the learned dictionary. }
    \label{label_fig1}
\end{figure}

In this study, we adopted the hypothesis that the PPG signal has a sparse structure. In particular, we assumed that the PPG signal can be decomposed using a few components, each describing a specific physiological event occurring in the signal, such as the systolic peak, the dicrotic notch, and the diastolic peak. Each of these components is zero everywhere except at the instant in which the corresponding event occurs.

Based on this sparse decomposition prior, we described the PPG signal using a convolutional sparse coding model \parencite{lewicki1998coding} \parencite{zeiler2010deconvolutional} \parencite{kavukcuoglu2010learning}. The main idea is that, given a dictionary of kernels $D$, the PPG signal is encoded into a sparse representation $X$ that collects the temporal activations of the kernels. The PPG signal is modelled as a sum of convolutions between each kernel and its temporal activation. Mathematically, the model can be expressed as:
\begin{subequations} \label{label_eq1}
    \begin{alignat}{1} 
    &\hat{y} = D \star X \,,\\
    &\text{with} \quad D \star X = \sum_{i=1}^{M}d_i * x_i\,, \quad \lVert X \rVert_0 \ll N \cdot M \,, \label{label_eq1b}
    \end{alignat}
\end{subequations}
where $\hat{y} \in \mathbb{R}^N$ is the approximated PPG signal, 
while the dictionary $D \in \mathbb{R}^{L \times M}$ collects $M$ kernels $d_i \in \mathbb{R}^L$. The matrix $X \in \mathbb{R}^{N \times M}$ collects the kernels' temporal activations $x_i \in \mathbb{R}^N$. $X$ is often called the sparse code. Indeed, the condition in equation \ref{label_eq1b} indicates that $X$ is sparse, where the $l_0$-norm $\lVert \cdot \rVert_0$ denotes the number of nonzero elements. Notice that the sparsity property has a twofold effect: $X$ should have a few nonzero columns, meaning that only a few kernels from $D$ are selected; and the nonzero columns should have only a few nonzero elements, meaning that the kernels are active only when the physiological events occur while silent elsewhere. The convolution operation $*$ ensures the shift-invariance property, since the same kernels can be reused at different time locations in the signal. Otherwise, the dictionary would have to include multiple shifted copies of the same kernels to account for all possible time locations \parencite{kavukcuoglu2010learning}. Moreover, since the kernels are shifted along the signal, the whole PPG signal can be processed without the need for segmentation of cardiac pulses. 

In our method, we adopted the design paradigm of algorithm unfolding. The core idea of unfolding is implementing each iteration of a model-based iterative algorithm using a single neural network layer, called the fold, and then stacking multiple of these layers together to execute a certain number of iterations \parencite{monga2021algorithm}. With this approach, the prior knowledge about the signal structure, namely the sparse decomposition model (equation \ref{label_eq1}), is embedded in the network architecture. In addition, dictionary learning on the dictionary $D$ is performed. The use of learned dictionaries has been shown to yield improved signal reconstruction and, in some cases, sparser representations \parencite{dumitrescu2018dictionary} \parencite{mairal2011task}. Furthermore, the learned kernels can be related to repeating patterns that might be present in the data \parencite{staerman2024unmixing}. 

Figure \ref{label_fig1} outlines the architecture of the proposed method. First, an encoder maps the noisy PPG signal $y_n$ into its sparse code $X_K$ given the current estimate of the dictionary $D$. This was implemented using the deep unfolded iterative shrinkage algorithm by \textcite{van2019deep}, which unfolds $K$ iterations of the iterative shrinkage-thresholding algorithm \parencite{daubechies2004iterative} and uses $2K-1$ trainable convolutional layers $W_{1,k}$ and $W_{2,k}$:
\begin{equation} \label{label_eq2}
    X_{k+1} = \mathcal{T}_{\theta_k}\left(W_{1,k}* y_n + W_{2,k} * X_k)\right)\,, 
\end{equation}
with $y_n$ the noisy PPG signal, $W_{1,k} \in \mathbb{R}^{L \times 1 \times M}$, $W_{2,k} \in \mathbb{R}^{L \times M \times M}$ and $\mathcal{T}_{\theta_k}$ the smooth soft-thresholding function \parencite{zhang2001thresholding}. The learned thresholds $\theta_k \in \mathbb{R}^{M}$ pass through a Softplus function before being used in the soft-thresholding function to avoid negative thresholds. Finally, the sparse code $X_K$ is fed to a decoder that reconstructs the signal and performs the dictionary learning task. As proposed by \textcite{sreter2018learned}, the decoder is a single-convolutional layer $D \in \mathbb{R}^{L \times M \times 1}$ that implements equation \ref{label_eq1}.

The method was trained by minimizing the following loss:
\begin{equation} \label{label_eq3}
    \argmin_{X_K,D} \frac{1}{2} \lVert y- \sum_{i=1}^{M}d_i * x_i \rVert^2_2 + \lambda \lVert X_K \rVert_1  \quad \text{s.t.} \quad \lVert d_i \rVert_2 = 1 \quad \forall \quad d_i \in D\,,
\end{equation}
where $y$ is the clean PPG signal and the $l_1$-norm regularization is applied on $X_K$ to promote sparsity. We trained the method to perform the denoising task in a supervised manner by feeding the artifact-corrupted PPG signals as input and using the clean PPG signals as target. The kernels were constrained to have unit norms; otherwise, $X_K$ would have vanished while the norm of the kernels would have exploded \parencite{moreau2017convolutional}.

We set the number of kernels to $M=32$, and their length $L=50$, which with $fs=125$ Hz corresponds to 0.4 seconds, based on the assumption that this duration is typically shorter than a full cardiac pulse, allowing each kernel to capture only local features within the pulses. The deep unfolded iterative shrinkage encoder was implemented using $K=10$ folds. We initialized the dictionary $D$ with random white noise. The $l_1$ regularization parameter $\lambda$ was set to 0.05, while $l_2$ regularization with parameter value $10^{-3}$ was applied to each $W_{1,k}$ and $W_{2,k}$ to avoid overfitting. We used the Adam optimizer with a learning rate of $10^{-4}$, and the batch size was set to 256. Early stopping occurred if the loss on the validation set was not improving for 10 consecutive epochs.

\begin{figure}[h!]
    \centering
    \includegraphics[width=\linewidth]{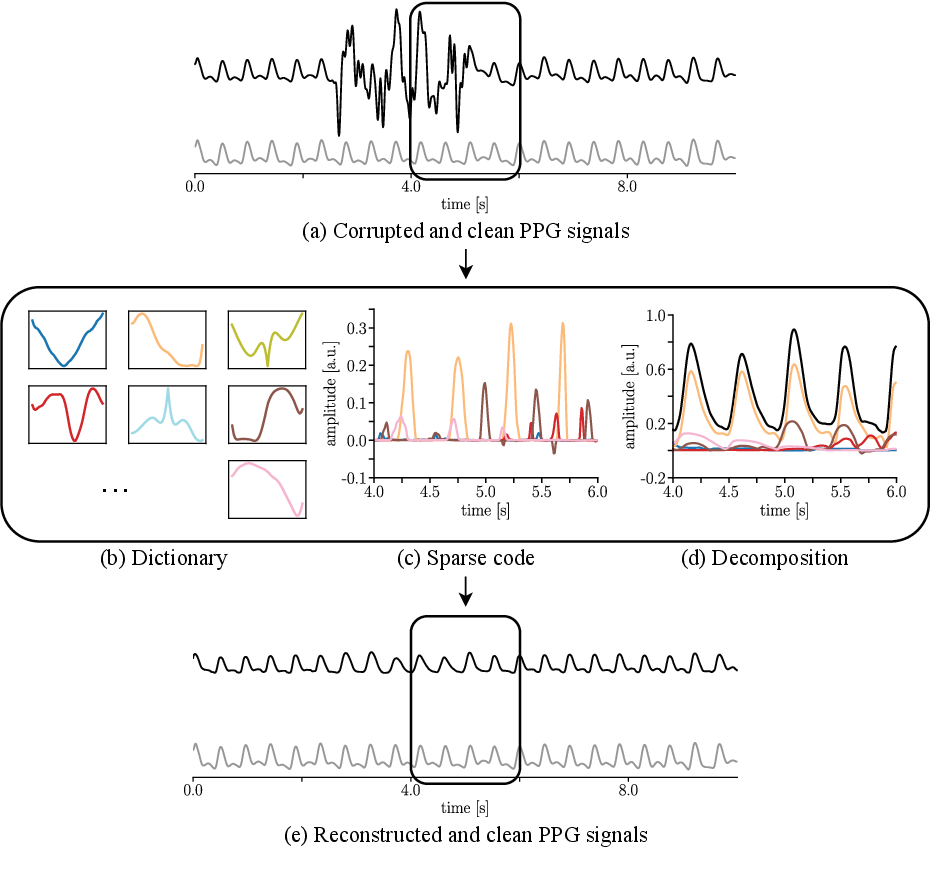}
    \caption{Illustration summarizing the core principles of the proposed method. (a) An example from the synthetic dataset with a corrupted PPG segment in black and its corresponding clean PPG segment in gray. A 2-seccond window is highlighted for illustration purposes. (b) A subset of the learned kernels from the dictionary. (c) Sparse code of the 2-second window of the corrupted PPG segment. The temporal activations of the kernels are depicted using the same color as in the dictionary. (d) Decomposition of the 2-second window. Each colored signal represents the convolution between a kernel and its temporal activation, while the black signal represents the reconstructed PPG signal obtained by summing the results of the convolutions (equation \ref{label_eq1}). (e) Corresponding reconstructed and clean 10-second PPG segments in black and gray, respectively.}
    \label{label_fig2}
\end{figure}

The core principles of the proposed method are summarized in Figure \ref{label_fig2}. The method takes as input the corrupted PPG signal (figure \ref{label_fig2}a) and reconstructs the signal using a dictionary of learned kernels (figure \ref{label_fig2}b). In particular, it encodes the input signal into its sparse code (figure \ref{label_fig2}c). Each column of the sparse code is the temporal activation of one kernel. Hence, the non-zero entries of the sparse code indicate which kernels are selected from the dictionary and at which time instants they are active. By convolving each column of the sparse code with the corresponding kernel, the signal's decomposition is computed (figure \ref{label_fig2}d). Finally, by summing these components, the denoised PPG is retrieved (figure \ref{label_fig2}e), as we expect that the kernels learned by the network are distinctive of the clean PPG signal's morphology.

\subsection{Reference methods}

We compared our proposed method with the CDA method by \textcite{mohagheghian2023noise} and the FC-GAN method by \textcite{avila2025generative} because of their strong performance and the availability of sufficient implementation details to ensure reproducibility.

In particular, the CDA method takes as input the STFT of the artifact-corrupted PPG signal and reconstructs the STFT of the clean PPG signal. Our training data had a sampling rate of 125 Hz, which was substantially higher than the 32 Hz used in \textcite{mohagheghian2023noise}. To maintain a comparable balance between model complexity and data resolution, we downsampled the synthetic dataset when training the CDA method. After testing various sampling rates, we selected 50 Hz as it yielded the best performance. The CDA method was trained using the STFT of the 10-second segments from the synthetic dataset. The signals were transformed using 6-seconds Hanning windows with 75\% overlap and with a number of frequency samples $\mathrm{nfft}=512$. Finally, the magnitude and phase of the transformed signals were resampled to $256 \times 32$ to match a power of 2, which simplified the implementation of the model's architecture.

The FC-GAN method uses an autoencoder as the generator, an encoder as the discriminator and an additional feature encoder. The latter learns to minimize the difference between the features extracted from the generated signals and the real signals. We implemented the three encoders and the decoder using four fully connected layers, as indicated in \parencite{avila2025generative}. Since in \parencite{avila2025generative} they did not report the number of units of each layer, we chose 1250, 600, 300, and 50 units for each encoder, while the number of units of the decoder mirrored that of the encoder. The method was trained using the loss indicated in \parencite{avila2025generative}, with all weights equal to 1.

Both CDA and FC-GAN methods were trained using the same optimizer settings and early stopping rule as our proposed method (see section \ref{section_2.2}).

\subsection{Evaluation criteria}
We first evaluated the performance of the methods on the synthetic dataset. Since the signals were artificially corrupted with synthetic motion artifacts, the clean target PPG signals were available. Hence, it was possible to evaluate the performance in terms of noise reduction using the signal-to-noise ratio (SNR), given as
\begin{equation} \label{label_eq4}
    \mathrm{SNR} = 10 \log_{10} \left( \frac{ \sum_{i=1}^{N} y^2}{ \sum_{i=1}^{N} (\hat{y}-y)^2} \right) \,,
\end{equation}
where $y$ is the clean target PPG and $\hat{y}$ is the PPG reconstruction obtained with one of the methods. The performance was further assessed on the synthetic dataset by examining the accuracy of the HR estimated after applying the methods. To do that, the systolic peaks of the PPG signals were detected using the Python toolbox Neurokit2 \parencite{makowski2021neurokit2} and the average HR of each 10-second segment was computed. We assessed the HR estimation using the mean absolute error (MAE) between the HR measured from the $i$-th denoised PPG segment $\mathrm{HR}_{\hat{y_i}}$ and the ground-truth HR measured from the clean PPG segment $\mathrm{HR}_{y_i}$:
\begin{equation} \label{label_eq5}
    \mathrm{MAE}_{\mathrm{HR}} = \frac{1}{W} \sum_{i=1}^{W} \lvert \mathrm{HR}_{\hat{y_i}} - \mathrm{HR}_{y_i} \rvert \,,
\end{equation}
where $W$ is the number of segments per subject and the HRs are given in beats-per-minute (bpm). Since the synthetic artifact model by \textcite{paliakaite2021modeling} gives the possibility of simulating four types of artifacts (see section \ref{label_sec2.1}), we analyzed the performance of all the methods with respect to each type of artifact. This was done by computing the boxplots of the $\mathrm{SNR}$ and $\mathrm{MAE}_{\mathrm{HR}}$ grouped by the artifact type. Since each subject was affected by only one artifact type, the averaging over segments within a subject in equation \ref{label_eq5} did not mix artifact types, making it possible to group the resulting $\mathrm{MAE}_{\mathrm{HR}}$ values by artifact type. We also grouped the boxplots by artifact duration, using 2-second-wide bins, to analyze the impact of artifact duration on the performance. Note that each subject had artifacts with random durations. Therefore, when grouping by artifact duration, the averaging in equation \ref{label_eq5} was performed over the subset of segments whose artifact durations fell within the same bin, resulting in one $\mathrm{MAE}_{\mathrm{HR}}$ value per subject for each artifact duration bin. Statistically significant differences in the $\mathrm{SNR}$ and $\mathrm{MAE}_{\mathrm{HR}}$ between the proposed method and each reference method (CDA method, FC-GAN method, as well as the metrics before denoising) were evaluated using the matched-pairs Wilcoxon signed-rank test. We tested both alternative hypotheses, namely that the proposed method had significantly better metrics than the reference methods, and vice versa. Only for the proposed method, the estimated HR was further assessed using a Bland-Altman plot, with mean and difference defined as
\begin{equation} \label{label_eq6}
    \overline{\mathrm{HR_i}} = \frac{1}{2}\left(\mathrm{HR}_{y_i} + \mathrm{HR}_{\hat{y_i}}  \right)\,,
\end{equation} 
and 
\begin{equation} \label{label_eq7}
    \Delta \mathrm{HR_i} =  \mathrm{HR}_{y_i} - \mathrm{HR}_{\hat{y_i}} \,.
\end{equation}
The limits of agreement were defined as the mean of $\Delta \mathrm{HR_i} \pm 1.96$ standard deviation of $\Delta \mathrm{HR_i}$, and a regression line was fitted.

Afterwards, the methods were tested on the PPG-DaLiA dataset. To apply the methods to the PPG-DaLiA dataset, we ensured that the input dimensions matched those used for training the methods on the synthetic dataset.  Since the methods were trained on 10-second segments, we used a sliding window approach with 10-second windows and a 2.5-second step, averaging the overlapping reconstructions. The performance of the methods were evaluated on PPG-DaLiA using the $\mathrm{MAE}_{\mathrm{HR}}$, where the ground-truth HR measured from ECG was provided in the dataset. Following the same approach used for the ground-truth HR \parencite{reiss2019deep}, the average HR was estimated over 8-second sliding windows with a 2-second step. Since during PPG-DaLiA recordings the subjects performed different types of daily activity, we assessed the performance of all methods with respect to each type of activity. This was done by grouping the boxplots of the $\mathrm{MAE}_{\mathrm{HR}}$ with the activity types. Following the same procedure as for the synthetic-agumented dataset, statistically significant differences in the boxplots were evaluated, and finally, the Bland-Altman plot of the HR estimation was computed.

\section{Results} \label{sec_results}

\begin{table}[h!]
\captionof{table}{$\mathrm{SNR}$ and $\mathrm{MAE}_{\mathrm{HR}}$ for the synthetic dataset, and $\mathrm{MAE}_{\mathrm{HR}}$ for the PPG-DaLiA dataset. Results are reported before denoising and after applying the CDA, the FC-GAN, and the proposed method. The numerical values are expressed as mean $\pm$ standard deviation.}
\begin{tabularx}{\textwidth}{
  l
  >{\centering\arraybackslash}X
  >{\centering\arraybackslash}X
  p{1.5em}  
  >{\centering\arraybackslash}X
}
\toprule
 & \multicolumn{2}{c}{Synthetic dataset} & & \multicolumn{1}{c}{PPG-DaLiA} \\
 & $\mathrm{SNR}$ [db] & $\mathrm{MAE}_{\mathrm{HR}}$ [bpm] & & $\mathrm{MAE}_{\mathrm{HR}}$ [bpm] \\
\midrule
Before denoising & $-7.06 \pm 8.44$  & $12.48 \pm 5.44$ & & $11.29 \pm 4.39$ \\
CDA              & $7.39 \pm 4.38$   & $9.67 \pm 6.45$  & & $12.03 \pm 4.29$ \\
FC-GAN           & $10.00 \pm 4.39$  & $5.72 \pm 5.12$  & & $9.48 \pm 5.09$ \\
Proposed method  & $\mathbf{11.23 \pm 5.39}$  & $\mathbf{5.62 \pm 3.46}$  & & $\mathbf{8.69 \pm 3.59}$ \\
\bottomrule
\end{tabularx}
\label{label_table1}
\end{table}

\begin{figure}[t!]
    \centering

    \begin{subfigure}{0.65\textwidth}
        \centering
        \includegraphics[width=\textwidth]{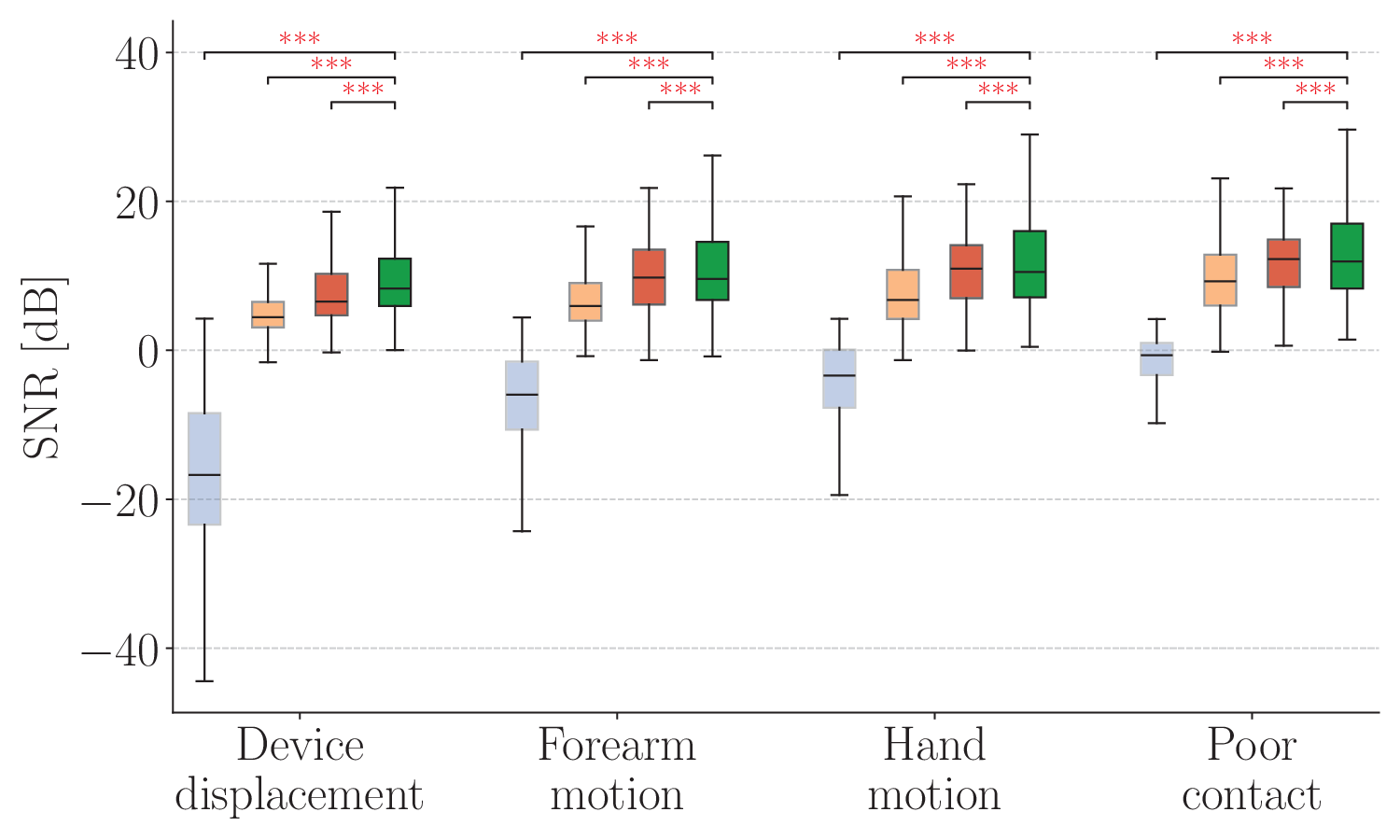}
        \caption{$\mathrm{SNR}$ grouped by artifact types}
        \label{label_fig3a}
    \end{subfigure}

    \vspace{1em}

    \begin{subfigure}{0.65\textwidth}
        \centering
        \includegraphics[width=\textwidth]{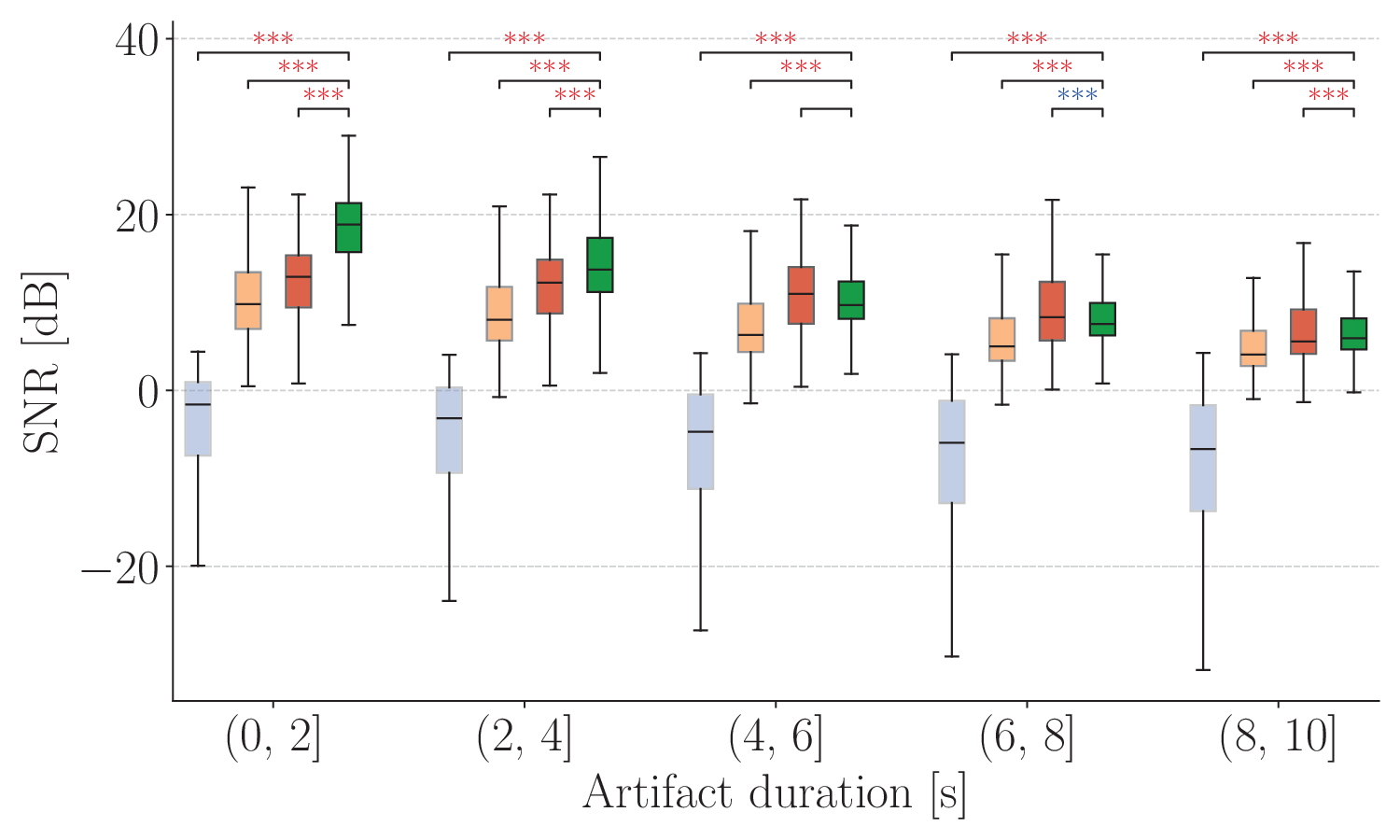}
        \caption{$\mathrm{SNR}$ grouped by artifact durations}
        \label{label_fig3b}
    \end{subfigure}

    \vspace{1em}

    \begin{subfigure}{0.5\textwidth}
        \centering
        \includegraphics[width=\textwidth]{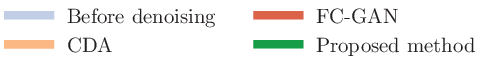}
    \end{subfigure}

    \caption{Boxplots of the $\mathrm{SNR}$ in the synthetic dataset. The graphs show the $\mathrm{SNR}$ before denoising and after applying the CDA, the FC-GAN and the proposed method. The boxplots are grouped by artifact types (figure \ref{label_fig3a}) and durations (figure \ref{label_fig3b}). The outcomes of the statistical test are reported with *** for $\text{p}<0.001$, ** for $\text{p}<0.01$ and * for $\text{p}<0.05$. Red asterisks indicate significantly higher $\mathrm{SNR}$ of the proposed method, while blue asterisks indicate significantly higher $\mathrm{SNR}$ of the reference method.}
    \label{label_fig3}
\end{figure}

The proposed and reference methods were first assessed on the synthetic dataset using the $\mathrm{SNR}$. Table \ref{label_table1} reports the mean and standard deviation of the evaluation metrics before denoising and after applying the CDA, FC-GAN and the proposed method. The $\mathrm{SNR}$ of the proposed method, on average, was higher than that of the signals before denoising and after applying the CDA and FC-GAN models. Figure \ref{label_fig3a} reports the boxplots of the $\mathrm{SNR}$ grouped by artifact types. Instead, in figure \ref{label_fig3b}, the boxplots are grouped based on artifact duration ranges of (0, 2], (2, 4], (4, 6], (6, 8], and (8, 10] seconds. Focusing on the $\mathrm{SNR}$ before applying the denoising, device displacement was the type resulting in the smallest $\mathrm{SNR}$ (figure \ref{label_fig3a}). Longer artifacts resulted in smaller $\mathrm{SNR}$ values in the signals before denoising (figure \ref{label_fig3b}). For all the artifact types and durations, the $\mathrm{SNR}$ of the proposed method was significantly higher than the $\mathrm{SNR}$ before denoising ($\text{p}<0.001$). Comparing the proposed and reference methods with different artifact types (figure \ref{label_fig3a}), the statistical test indicated that the proposed method had significantly larger $\mathrm{SNR}$ than both reference methods for all the types, with $\text{p}<0.001$. We can also notice that the proposed method had larger standard deviations than the reference methods. Considering the performance with different durations (figure \ref{label_fig3b}), there is statistically significant evidence that the proposed method had a larger $\mathrm{SNR}$ than the CDA method for all the duration ranges, with $\text{p}<0.001$. Moreover, the proposed method had significantly larger $\mathrm{SNR}$ than the FC-GAN method for the ranges (0, 2], (2, 4] and (8, 10] seconds ($\text{p}<0.001$). Instead, for the range (4, 6] seconds, there were no statistically significant differences, while for the range (6, 8] seconds, the FC-GAN had a significantly higher $\mathrm{SNR}$ ($\text{p}<0.001$).

\begin{figure}[t!]
    \centering

    \begin{subfigure}{0.65\textwidth}
        \centering
        \includegraphics[width=\textwidth]{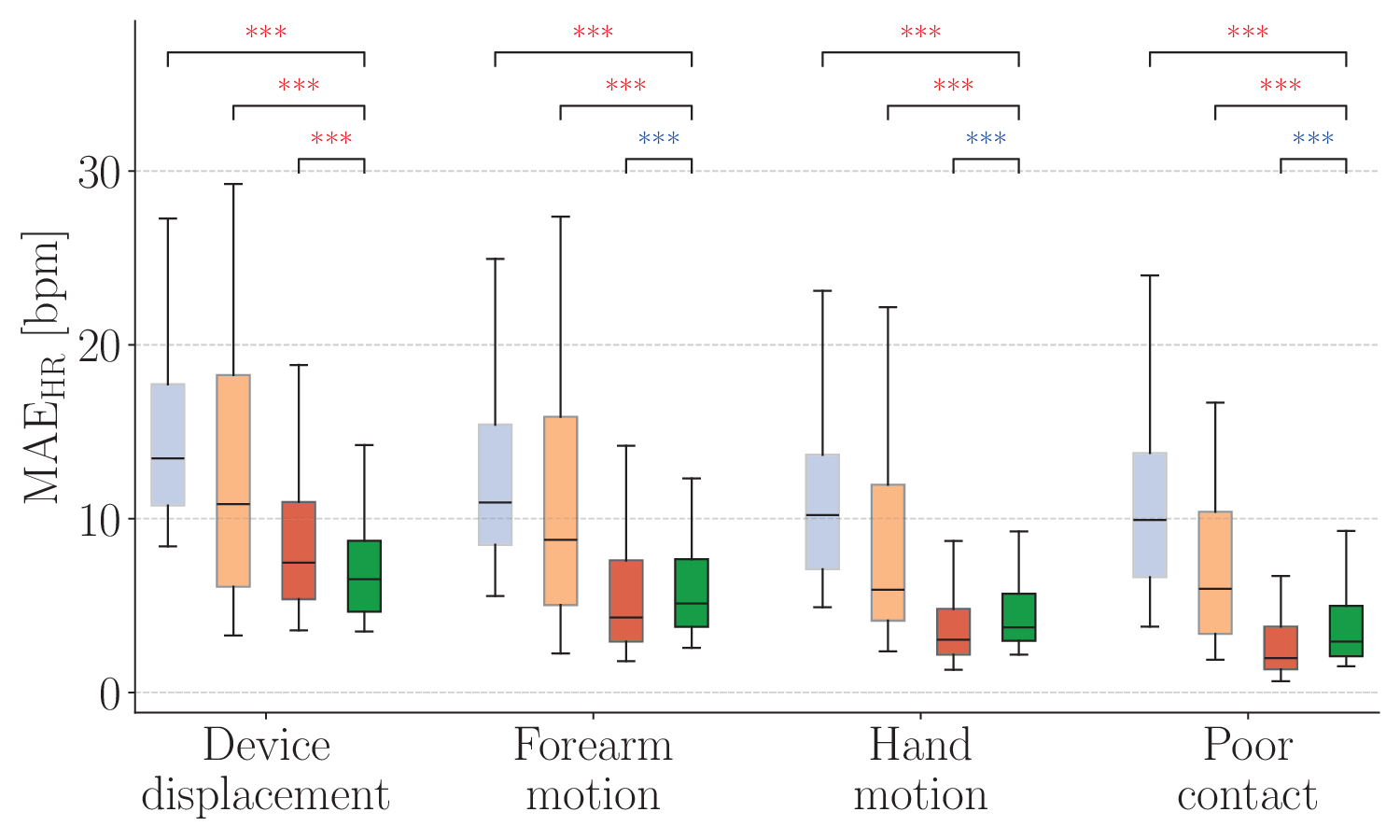}
        \caption{$\mathrm{MAE}_{\mathrm{HR}}$ grouped by artifact types}
        \label{label_fig4a}
    \end{subfigure}

    \vspace{1em}

    \begin{subfigure}{0.65\textwidth}
        \centering
        \includegraphics[width=\textwidth]{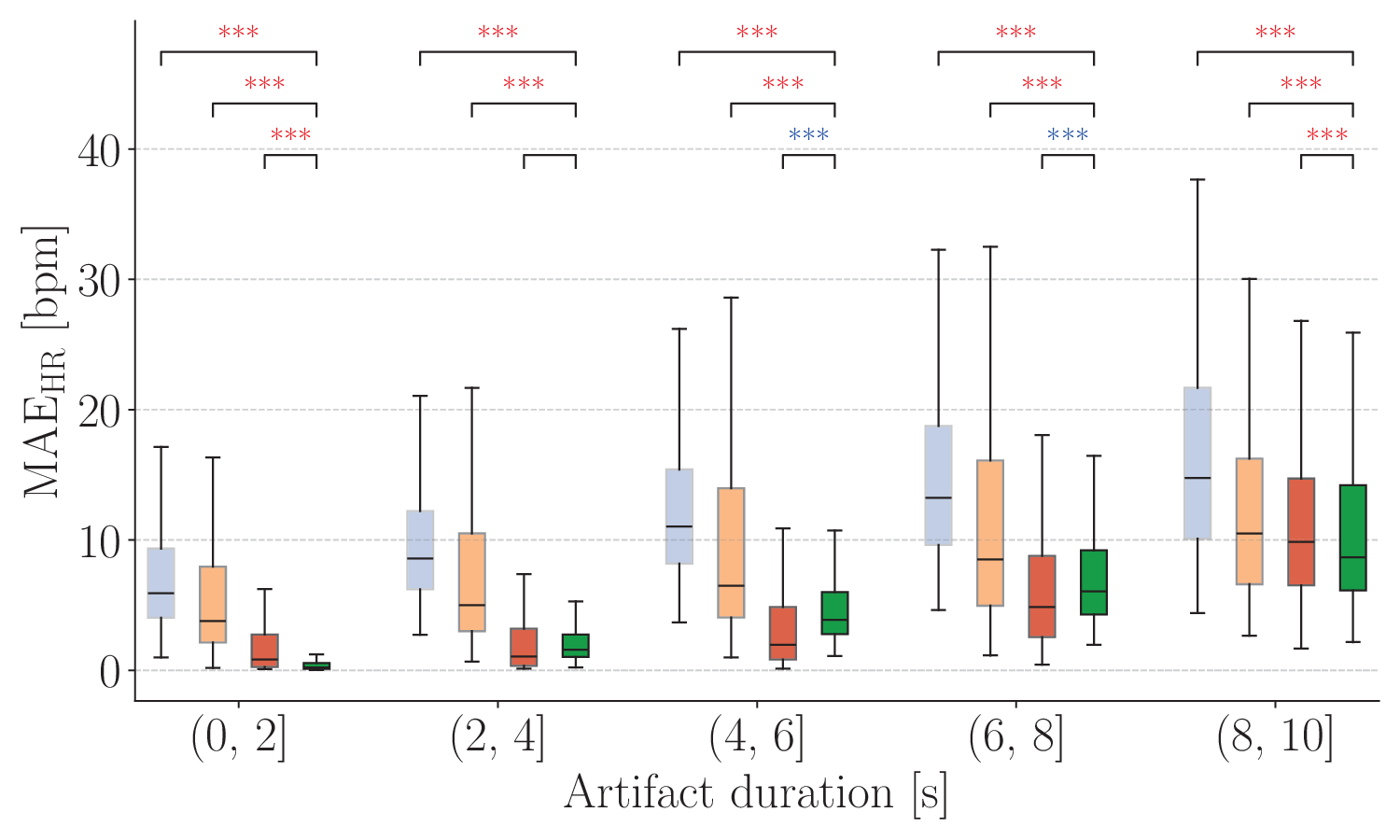}
        \caption{$\mathrm{MAE}_{\mathrm{HR}}$ grouped by artifact durations}
        \label{label_fig4b}
    \end{subfigure}

    \vspace{1em}

    \begin{subfigure}{0.5\textwidth}
        \centering
        \includegraphics[width=\textwidth]{legend.eps}
    \end{subfigure}

    \caption{Boxplots of the $\mathrm{MAE}_{\mathrm{HR}}$ in the synthetic dataset. The graphs show the $\mathrm{MAE}_{\mathrm{HR}}$ before denoising and after applying the CDA, the FC-GAN and the proposed method. The boxplots are grouped by artifact types (figure \ref{label_fig4a}) and durations (figure \ref{label_fig4b}). The outcomes of the statistical test are reported with *** for $\text{p}<0.001$, ** for $\text{p}<0.01$ and * for $\text{p}<0.05$. Red asterisks indicate significantly smaller $\mathrm{MAE}_{\mathrm{HR}}$ of the proposed method, while blue asterisks indicate significantly smaller $\mathrm{MAE}_{\mathrm{HR}}$ of the reference method.}
    \label{label_fig4}
\end{figure}

To further evaluate the performance of the algorithms on the synthetic dataset, we assessed the HR estimated after applying each algorithm. As reported in table \ref{label_table1}, the $\mathrm{MAE}_{\mathrm{HR}}$ of the proposed method was, on average, 55\% lower than that of the signals before denoising. The proposed method also showed a lower mean $\mathrm{MAE}_{\mathrm{HR}}$ compared to both reference methods. Furthermore, the $\mathrm{MAE}_{\mathrm{HR}}$ of the proposed method had a smaller standard deviation than the reference methods. Figure \ref{label_fig4a} shows the boxplots of the $\mathrm{MAE}_{\mathrm{HR}}$ grouped by artifact types, while figure \ref{label_fig4b} shows the boxplots of the $\mathrm{MAE}_{\mathrm{HR}}$ grouped by artifact durations. In figure \ref{label_fig4a}, the device displacement artifact resulted in the largest $\mathrm{MAE}_{\mathrm{HR}}$ before denoising. In figure \ref{label_fig4b}, longer artifacts yielded larger $\mathrm{MAE}_{\mathrm{HR}}$. The proposed method had a significantly smaller $\mathrm{MAE}_{\mathrm{HR}}$ than before denoising and the CDA method for all types and durations of artifacts ($\text{p}<0.001$). Comparing the proposed method with the FC-GAN method in figure \ref{label_fig4a}, the proposed method had a significantly smaller $\mathrm{MAE}_{\mathrm{HR}}$ for device displacement type ($\text{p}<0.001$), while the FC-GAN method had a significantly smaller $\mathrm{MAE}_{\mathrm{HR}}$ for the other 3 types of artifacts ($\text{p}<0.001$). With regard to the artifact durations (figure \ref{label_fig4b}), the proposed method had a significantly smaller $\mathrm{MAE}_{\mathrm{HR}}$ than the FC-GAN method for the ranges (0, 2], (8, 10] seconds ($\text{p}<0.001$), the FC-GAN method had a significantly smaller $\mathrm{MAE}_{\mathrm{HR}}$ for the ranges (4, 6] and (6, 8] seconds ($\text{p}<0.001$), while there were no statistically significant differences for the range (2, 4] seconds.

\begin{figure}[t!]
    \centering

    \begin{subfigure}{0.9\textwidth}
        \centering
        \includegraphics[width=\textwidth]{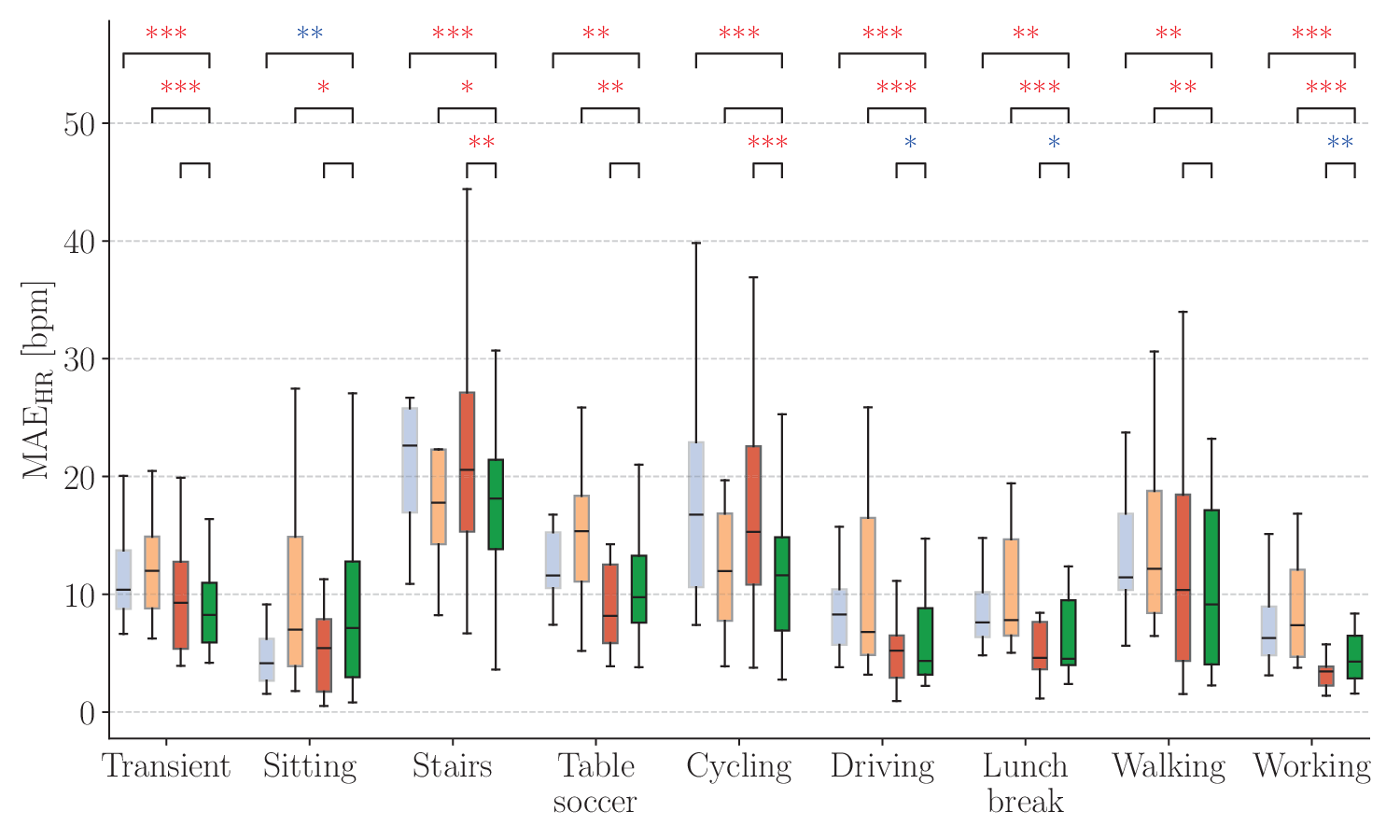}
        
        \vspace{0.5em}
        {\small{$\mathrm{MAE}_{\mathrm{HR}}$ grouped by activity types}}
    \end{subfigure}

    \vspace{1em}

    \begin{subfigure}{0.5\textwidth}
        \centering
        \includegraphics[width=\textwidth]{legend.eps}
    \end{subfigure}

    \caption{Boxplots of the $\mathrm{MAE}_{\mathrm{HR}}$ in the PPG-DaLiA dataset. The graph shows the $\mathrm{MAE}_{\mathrm{HR}}$ before denoising and after applying the CDA, the FC-GAN and the proposed method. The boxplots are grouped by activity types. The outcomes of the statistical test are reported with *** for $\text{p}<0.001$, ** for $\text{p}<0.01$ and * for $\text{p}<0.05$. Red asterisks indicate significantly smaller $\mathrm{MAE}_{\mathrm{HR}}$ of the proposed method, while blue asterisks indicate significantly smaller $\mathrm{MAE}_{\mathrm{HR}}$ of the reference method.}
    \label{label_fig5}
\end{figure}
Then, the algorithms were tested on the PPG-DaLiA dataset. Regarding this, the $\mathrm{MAE}_{\mathrm{HR}}$ of the proposed method, on average, was 23\% smaller than that of the signals before denoising. Additionally, the proposed method had a lower mean $\mathrm{MAE}_{\mathrm{HR}}$ compared to the reference methods. Figure \ref{label_fig5} reports the boxplots of the $\mathrm{MAE}_{\mathrm{HR}}$ divided by the type of activity performed by the subjects. For the sitting activity, the signals before denoising had significantly smaller $\mathrm{MAE}_{\mathrm{HR}}$ than after applying the proposed method ($\text{p}<0.01$). For the remaining activities, the proposed method had significantly smaller $\mathrm{MAE}_{\mathrm{HR}}$ than before denoising with $\text{p}<0.001$ for 5 activities, while with $\text{p}<0.01$ for 3 activities. Comparing the proposed method with the CDA method, the proposed method had a significantly smaller $\mathrm{MAE}_{\mathrm{HR}}$ with $\text{p}<0.001$ for 4 activities, with $\text{p}<0.01$ for 2 activities, and with $\text{p}<0.05$ for 2 activities, while one activity had no statistically significant differences. Comparing the proposed method with the FC-GAN method, the proposed method had a significantly smaller $\mathrm{MAE}_{\mathrm{HR}}$  for two activities, one with $\text{p}<0.01$ and the other with $\text{p}<0.001$; the FC-GAN had smaller $\mathrm{MAE}_{\mathrm{HR}}$ with $\text{p}<0.01$ for one activity and $\text{p}<0.05$ for 2 activities, while for the remaining 4 activities there were no statistically significant differences.

\begin{figure}[t!]
    \centering

    \begin{subfigure}{0.48\textwidth}
        \centering
        \includegraphics[width=\textwidth]{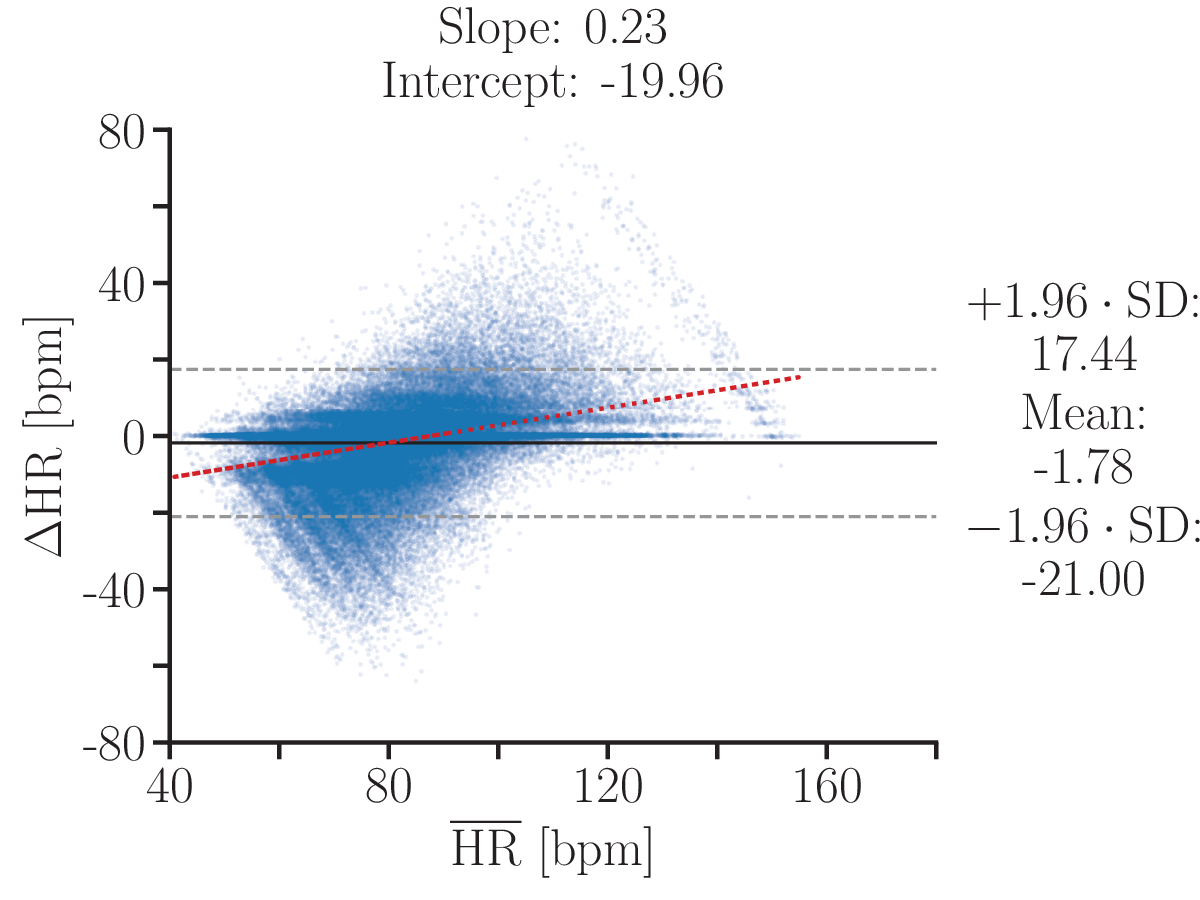}
        \caption{Synthetic dataset}
        \label{label_fig6a}
    \end{subfigure}
    \hfill
    \begin{subfigure}{0.48\textwidth}
        \centering
        \includegraphics[width=\textwidth]{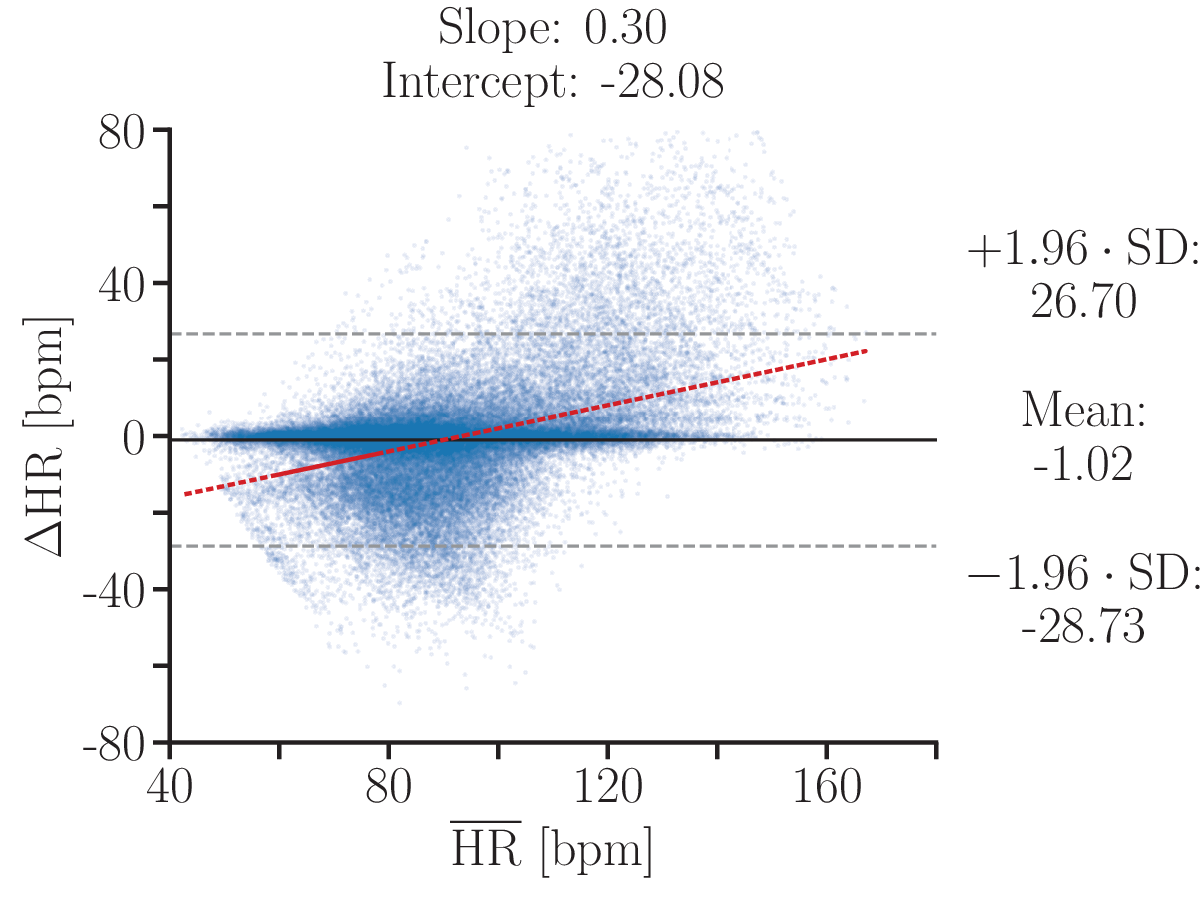}
        \caption{PPG-DaLiA}
        \label{label_fig6b}
    \end{subfigure}

    \caption{Bland-Altman plots of the HR estimated after applying the proposed method on the synthetic dataset (figure \ref{label_fig6a}) and PPG-DaLiA dataset (figure \ref{label_fig6b}). $\overline{\mathrm{HR_i}}$ and $\Delta \mathrm{HR}$ are the mean and difference between the ground-truth HR and the HR estimated from the reconstructed PPG (equations \ref{label_eq6} and \ref{label_eq7}). The mean of $\Delta \mathrm{HR}$ is reported with a black solid line, the limits of agreement $\Delta \mathrm{HR} \pm 1.96$ standard deviation of $\Delta \mathrm{HR}$ with grey dashed lines and the regression line with a red dotted line. Their numerical values are also included.}
    \label{label_fig6}
\end{figure}

Finally, the performance was further assessed using the Bland-Altman plot on both the synthetic and PPG-DaLiA datasets, as depicted in figures \ref{label_fig6a} and \ref{label_fig6b}, respectively. For both datasets, the regression lines had a positive slope, and the means of $\Delta \mathrm{HR}$ were negative. The limits of agreement on the PPG-DaLiA dataset were larger than the ones on the synthetic dataset. 

\section{Discussion} \label{sec_discussion}
The findings of our study suggested that the proposed method effectively improves the quality of PPG signals, reducing the impact of motion artifacts. This is supported by the consistently higher $\mathrm{SNR}$ and lower $\mathrm{MAE}_{\mathrm{HR}}$ observed after denoising in both datasets (Table \ref{label_table1}). Furthermore, on the synthetic dataset, the proposed method significantly improved the $\mathrm{SNR}$ and $\mathrm{MAE}_{\mathrm{HR}}$ compared to the signals before denoising for all the artifact types and durations (figures \ref{label_fig3} and \ref{label_fig4}). Similarly, in the PPG-DaLiA dataset, the proposed method reduced the $\mathrm{MAE}_{\mathrm{HR}}$ for all types of activity except sitting activity (figure \ref{label_fig5}). Interestingly, for this activity, not applying any denoiser yielded a more accurate HR estimate than either the reference methods or the proposed method. In the sitting condition, motion artifacts are likely minimal, resulting in relatively clean signals. Hence, the temporal features critical for HR estimation, such as peak timing and sharpness, are well preserved. Instead, applying any of the denoising methods may distort these features, leading to reduced accuracy in HR estimation compared to the unprocessed signals. In this situation, a signal quality analysis could be performed to decide whether the use of the denoiser is required.

The results showed that the proposed method outperformed the CDA method in terms of enhancing the signal quality and recovering precise temporal features necessary for accurate HR estimation. 
Overall, this is shown by the higher mean $\mathrm{SNR}$ and smaller mean $\mathrm{MAE}_{\mathrm{HR}}$ in table \ref{label_table1}. Furthermore, the proposed method led to superior values of $\mathrm{SNR}$ and $\mathrm{MAE}_{\mathrm{HR}}$ for all artifact types and durations in the synthetic dataset (figures \ref{label_fig3} and \ref{label_fig4}), as well as smaller values of $\mathrm{MAE}_{\mathrm{HR}}$ for 8 out of 9 activities in the PPG-DaLiA dataset (figure \ref{label_fig5}). The lower performance of the CDA method may be related to the use of the STFT, whose major limitation is the trade-off between time and frequency resolutions imposed by the window's length. Favouring a high frequency resolution at the cost of a lower time resolution can improve the HR estimation by preserving the dominant spectral components, but it may compromise temporal features, leading to distorted peaks' morphology or shifted peaks. Conversely, improved time resolution may better preserve the temporal features but results in poorer frequency resolution and, hence, HR estimation accuracy.

Our results indicate that the proposed method was more effective than the FC-GAN method at restoring the signals' quality. This is shown by the fact that the proposed method obtained a larger mean $\mathrm{SNR}$ (table \ref{label_table1}) and a significantly larger $\mathrm{SNR}$ for all the artifact types and for most of the durations (figures \ref{label_fig3} and \ref{label_fig4}). The higher signals' quality suggests that the proposed method may be more suitable than the FC-GAN method for applications where accurate reconstruction of the signal morphology is important, as, for example, applications in which the extraction of morphological features is used to estimate arterial stiffness or blood pressure. Instead, in terms of HR estimation, the proposed and FC-GAN methods had similar performance. On average, the proposed method achieved a slightly smaller $\mathrm{MAE}_{\mathrm{HR}}$ on the synthetic dataset and smaller $\mathrm{MAE}_{\mathrm{HR}}$ on the PPG-DaLiA dataset (table \ref{label_table1}). Interestingly, the proposed method obtained lower $\mathrm{MAE}_{\mathrm{HR}}$ only for signals with the most severe motion artifacts. On the synthetic dataset, this is the case of the device displacement type. Device displacement causes the largest artifact amplitude \parencite{paliakaite2021modeling} and, accordingly, it was the type that had smaller $\mathrm{SNR}$ and larger $\mathrm{MAE}_{\mathrm{HR}}$ in the signals before denoising. Similarly, on the PPG-DaLiA dataset, the proposed method achieved lower $\mathrm{MAE}_{\mathrm{HR}}$ than the FC-GAN method for the activities ascending/descending stairs and cycling. These activities introduced the most severe artifacts and were associated with the poorest performance before denoising. Furthermore, in most cases, the proposed method achieved lower standard deviations in the $\mathrm{MAE}_{\mathrm{HR}}$ than the FC-GAN method (table \ref{label_table1} and figure \ref{label_fig4}), indicating higher consistency in the HR measurement.

For both datasets, the Bland–Altman plots showed a small negative mean $\Delta \mathrm{HR}$, indicating a fixed bias. Hence, the HR measured after applying the proposed method was slightly overestimated. Moreover, the regression line highlighted the presence of a positive proportional bias for both datasets, suggesting that greater overestimations occurred at lower HR values. The limits of agreement in the PPG-DaLiA dataset were wider than in the synthetic dataset, indicating greater uncertainty in HR measurements. This was expected since the PPG-DaLiA dataset includes real corrupted signals during various daily life actions.

The key advantage of the proposed approach over the previous deep learning models is its interpretability, as it can extract meaningful waveform features. As shown in figure \ref{label_fig2}, the method learned a dictionary of kernels which seem to resemble some recurrent morphological patterns in the signals. Thanks to the prior convolutional sparse coding model, the temporal activations in the sparse code resemble some Dirac deltas. These deltas indicate at which time locations the patterns occur in the signal, and highlight the periodic characteristics of the PPG signal. Hence, the potential of the method might extend beyond denoising, as these features might be precious for developing novel diagnostic tools. For instance, the temporal activations in the sparse code might be used to detect critical pathological events. Moreover, if the kernels that characterize a subject’s signal morphology differ noticeably from those typically observed in healthy individuals, this could help to detect a potential cardiovascular disease. The interpretability of the model could be further improved by tying the weights of the encoder and decoder, since in \parencite{tolooshams2018scalable} they demonstrated that this strategy can learn a ground-truth dictionary.

It is worth noticing that we did not compare the performance with state-of-the-art methods specifically designed for HR estimation, e.g. \parencite{zhang2014troika} \parencite{reiss2019deep}, because this was not the main purpose of our method. Indeed, we used a publicly available algorithm to estimate HR from the denoised signals only to evaluate the denoising performance.

The outcomes of this study should be interpreted in light of certain limitations, mainly associated with the data available to train the method. In practice, obtaining paired corrupted PPG signals and their corresponding clean reference signals is not feasible \parencite{jain2023self}. As a result, training neural networks for PPG denoising typically relies either on synthetic artifact generation, as done in our study, or on protocols designed to induce artifacts in a controllable manner. For example, some studies recorded PPG signals simultaneously from both hands, with one hand still and the other performing specific movements \parencite{roy2018improving} \parencite{xu2019photoplethysmography}, and they considered the signals from the moving hand as noisy signals while the ones from the still hand as the clean target signals. Both approaches have inherent limitations. Synthetic artifacts may not fully capture the complexity and variability of motion artifacts encountered in real-life scenarios. The protocols with simultaneous two-hand recordings assume that the PPG signals from both hands are identical, which may not hold in practice. Moreover, the range of movements that can be introduced in such controlled settings is limited compared to the variability of natural movements. Another limitation of our study is associated with the fact that the method was trained using fingertip PPG,
even though we aimed to apply the method to wrist PPG, which is commonly used in wearable devices. Evidently, there are no large publicly available datasets containing clean wrist PPG signals, as wrist recordings are usually done in daily life. In contrast, clean fingertip PPG datasets are more common, since fingertip measurements are often acquired in hospitals from bedside patients. Our training data may have introduced some shortcomings because the morphology of finger PPG might be considerably different from that of wrist PPG. Furthermore, the range of HR associated with bedside patients is not as wide as for subjects doing daily activities. Nonetheless, the proposed method showed promising performance also with wrist PPG signals from the PPG-DaLiA dataset.

\section{Conclusion} \label{sec_conclusion}
In this study, we introduced a novel deep-learning method for denoising PPG signals. Our objective was to combine the advantages of both traditional signal decomposition methods and modern deep-learning approaches. Specifically, we aimed to create a technique with good interpretability while achieving the enhanced performance of deep learning approaches. This was achieved by integrating prior knowledge about the sparse structure of the PPG signal into the network architecture using algorithm unfolding.

Testing the proposed method on a dataset with synthetic motion artifacts allowed us to assess the performance with respect to the ground-truth signal, under various conditions of artifact types and durations. The method was then tested using real signals from a wearable device during daily life activities, allowing us to evaluate its performance in a more realistic scenario. The comparison of the signal-to-noise ratio and the mean absolute error of the HR estimation before and after applying the proposed method demonstrated that our approach can successfully enhance the signal quality and the accuracy of the HR measurements. The comparison of the proposed method with the reference CDA and FC-GAN methods suggested that our method is more effective in improving signal quality and can achieve accuracy in HR estimation comparable to that of the reference methods.

The next step of our research could be investigating whether the features encoded in the sparse code and in the dictionary can highlight some patterns associated with the presence of cardiovascular diseases. Hence, the potential of the method extends beyond denoising, since these features may reveal ongoing cardiovascular alterations and open the avenues for innovative diagnostic tools.

\section*{Data availability statement}
This research used the data from the PulseDB dataset \parencite{wang2023pulsedb}, which is available at \url{https://github.com/pulselabteam/PulseDB} and from PPG-DaLiA dataset \parencite{reiss2019deep} which is available at \url{https://ubicomp.eti.uni-siegen.de/home/datasets/sensors19/}. No new data were created or analyzed in this study. 

\ack
This work was labelled by ITEA and funded by local authorities under grant agreement "ITEA-2021-21022-RM4Health". Furthermore, the research was performed within the framework of the Eindhoven MedTech Innovation Center (e/MTIC, incorporating Eindhoven University of Technology, Royal Philips, Catharina Hospital, Maxima Medical Center and Kempenhaeghe Epilepsy and Sleep Center).

\section*{ORCID iD}
Giulio Basso \orcidlink{0000-0002-4532-4354} \url{ https://orcid.org/0000-0002-4532-4354}.\\
Xi Long \orcidlink{0000-0001-9505-1270} \url{https://orcid.org/0000-0001-9505-1270}.\\
Reinder Haakma \orcidlink{0000-0002-7245-7072} \url{ https://orcid.org/0000-0002-7245-7072}.\\
Rik Vullings \orcidlink{0000-0002-2392-6098} \url{ https://orcid.org/0000-0002-2392-6098}.

\printbibliography

\end{document}